\documentclass[a4paper,11pt]{article}
\pdfoutput=1 
\usepackage{jheppub} 
\usepackage{epsfig}
\usepackage{caption}
\usepackage{yfonts}
\usepackage{subcaption}
\usepackage{color,soul}
\usepackage{float}
\usepackage{amsfonts}
\usepackage{url}
\usepackage{slashed}
\usepackage{accents}
\usepackage{bbm,multicol}
\usepackage{lipsum}
\usepackage{mathtools}
\usepackage{physics}

\usepackage[normalem]{ulem}

\hypersetup{
  bookmarks=true,         
  unicode=false,          
  pdftoolbar=true,        
 pdfmenubar=true,        
 pdffitwindow=true,     
 pdfstartview={FitH},    
 pdfsubject={Dark Matter},   
 pdfnewwindow=true,      
 pdfcreator={RevTeX},
 colorlinks=true,       
 linkcolor=red,          
 citecolor=blue,        
 filecolor=black,      
 urlcolor=blue,           
  }

\title{Revisiting the role of CP-conserving processes in cosmological particle-antiparticle asymmetries}
\author[]{Avirup Ghosh,}
\author[]{Deep Ghosh}
\author[]{and Satyanarayan Mukhopadhyay}
\affiliation[]{School of Physical Sciences, Indian Association for the Cultivation of Science, 2A and 2B Raja S.C. Mullick Road, Kolkata 700 032}

\emailAdd{avirup.ghosh1993@gmail.com}
\emailAdd{tpdg@iacs.res.in}
\emailAdd{tpsnm@iacs.res.in}

\abstract{We point out qualitatively different possibilities on the role of CP-conserving processes in generating cosmological particle-antiparticle asymmetries, with illustrative examples from models in leptogenesis and asymmetric dark matter production. In particular, we consider scenarios in which the CP-violating and CP-conserving processes are either both decays or both scatterings, thereby being naturally of comparable rates. This is in contrast to the previously considered CP-conserving processes in models of leptogenesis in different see-saw mechanisms, in which the CP-conserving scatterings typically have lower rates compared to the CP-violating decays, due to a Boltzmann suppression. We further point out that the CP-conserving processes can play a dual role if the asymmetry is generated in the mother sector itself, in contrast to the conventional scenarios in which it is generated in the daughter sector. This is because, the CP-conserving processes initially suppress the asymmetry generation by controlling the out-of-equilibrium number densities of the bath particles, but subsequently modify the ratio of particle anti-particle yields at the present epoch by eliminating the symmetric component of the bath particles through pair-annihilations, leading to a competing effect stemming from the same process at different epochs. We find that the asymmetric yields for relevant particle-antiparticle systems can vary by orders of magnitude depending upon the relative size of the CP-conserving and violating reaction rates.}

\begin{document} 
\maketitle
\flushbottom  
\section{Introduction}
\label{sec:sec1}
Cosmological production of particle-antiparticle asymmetries through dynamical mechanisms has been a topic of extensive studies~\cite{Weinberg:2008zzc}. In particular, several possible ways of generating the observed baryon-antibaryon asymmetry of the Universe (BAU) have been proposed~\cite{Yoshimura:1978ex,Ignatiev:1978uf, Weinberg:1979bt, Nanopoulos:1979gx, Yoshimura:1979gy, Affleck:1984fy, Kuzmin:1985mm, Fukugita:1986hr}. Starting from a symmetric initial condition with zero asymmetry, as implied by inflationary scenarios, achieving a non-zero BAU requires satisfying the three Sakharov conditions of baryon number (B) violation, the violation of charge conjugation (C) and charge conjugation - parity (CP) discrete symmetries, and departure from thermodynamic equilibrium~\cite{Sakharov:1967dj, Weinberg:2008zzc}. Two of the most well studied BAU generation mechanisms are baryogenesis in grand unified theories (GUT)~\cite{Yoshimura:1978ex,Weinberg:1979bt, Nanopoulos:1979gx, Yoshimura:1979gy}, and baryogenesis through leptogenesis~\cite{Fukugita:1986hr}. In both scenarios, most of the implementations considered involve the CP-violating out-of-equilibrium decay of a heavy particle in the early Universe. Thus, the primary quantities that determine the net asymmetry generated are the rates of the CP-violating decay and its inverse process. If in addition CP-violating scattering processes are present, they are found to affect both the generated asymmetry, and its subsequent wash-out, depending upon the hierarchy of the decoupling temperatures for the decay and scattering processes; for reviews, see, for example, Refs.~\cite{Kolb:1990vq, Buchmuller:2005eh, Davidson:2008bu}, and references therein. 

The primary source of the CP-violation in a given scenario may also turn out to be scattering processes, with the heavy particle decay processes either sub-dominant or absent~\cite{Bento:2001rc, Nardi:2007jp, Gu:2009yx}. Such scattering mechanisms have been studied in contexts involving the dark matter (DM) particles as well, in achieving baryogenesis through DM annihilations~\cite{Cui:2011ab, Bernal:2012gv, Bernal:2013bga, Kumar:2013uca},  in relating the baryon and dark matter sector asymmetries~\cite{Baldes:2014gca, Baldes:2015lka}, in realizing the asymmetric DM (ADM) scenario through scatterings~\cite{GGM_ADM_Z2}, and through semi-annihilations~\cite{Ghosh:2020lma, DEramo:2010keq}. 

The role of CP-conserving processes have also been studied in the context of leptogenesis in see-saw models of neutrino mass. In leptogenesis from type-I see-saw, CP-conserving scatterings can arise from additional new interactions of the SM gauge singlets~\cite{Gu:2009hn, Sierra:2014sta}. As discussed in Ref.~\cite{Sierra:2014sta}, these additional interactions are CP-conserving by assumption, that is in general they can violate CP if the relevant couplings have complex phases. Furthermore, it is a specific dependence of the reaction rates on the temperature and masses that leads to this competing effect, in spite of one process being a decay and the other a scattering, as analyzed in detail in~\cite{Sierra:2014sta}. In leptogenesis from type-II and type-III see-saw, there are additional gauge interactions, which lead to CP-conserving scatterings, see, for example~\cite{Hambye:2003rt, Hambye:2005tk, Hambye:2012fh}. The rate of these gauge scatterings are usually smaller than the CP-violating decay rates, and therefore as shown in Ref.~\cite{Hambye:2005tk}, the role of the gauge scatterings in determining the asymmetry is small, and the efficiency factor remains close to maximal.

In this paper, we revisit the role of CP-conserving reactions in determining cosmological particle-antiparticle asymmetries, and point out qualitatively novel scenarios in which their role is significantly exemplified. In particular, to begin with, we may classify the possible scenarios into three broad categories:
\begin{enumerate}
\item[\bf{A.}] both the CP-violating and CP-conserving processes are decays

\item[\bf{B.}] both the CP-violating and CP-conserving processes are scatterings

\item[\bf{C.}] one of the processes is a decay and the other is a scattering
\end{enumerate}
In scenarios A and B, the two processes can naturally have comparable rates, and therefore the CP-conserving reactions can in principle play a significant role in generating the particle-antiparticle asymmetries. In scenario C, which is the case so far considered in the literature, it is expected that the CP-conserving scatterings will have a suppressed reaction rate compared to the CP-violating decays, due to the Boltzmann suppression stemming from the presence of an additional non-relativistic particle in the initial state. In this paper, we shall discuss examples of both scenarios A and B. 

In each of these scenarios, one can further have two distinct possibilities. For processes of the form $M... \rightarrow D...$, where $M$ is heavier than $D$, we can have a scenario in which
\begin{enumerate}
\item[\bf{I.}] the asymmetry is generated in the mother sector, i.e., in the $M$ sector,

\item[\bf{II.}] the asymmetry is generated in the daughter sector, i.e., in the $D$ sector.
\end{enumerate}
In conventional scenarios of baryogenesis, we usually find examples of type II. In this paper, we shall show examples of both types, belonging to scenarios A and B above. In particular, we shall point out examples of type I, in which the CP-conserving processes play a dual role. Until the decoupling of the CP-violating scatterings, the CP-conserving processes tend to suppress the asymmetry generation by controlling the out-of-equilibrium number densities of the bath ($M$) particles. However, subsequently, once the net particle antiparticle yield difference has been frozen, the same CP-conserving pair-annihilation reactions modify the ratio of particle anti-particle yields by eliminating the symmetric component of the mother particles~\footnote{Since either the DM particle ($\chi$) or the DM anti-particle ($\chi^\dagger$) may dominate the current DM density, we quantify this ratio by the asymptotic value of the asymmetry parameter $r_\infty=\lvert Y_{\chi} - Y_{\chi^\dagger}\rvert / (Y_{\chi} + Y_{\chi^\dagger})$ in the present epoch, where $Y_i$ are the respective yields; see Sec.~\ref{sec:sec3} for details.}. This leads to a competing effect shown by the same scattering process at different epochs, since, unlike in the scenarios of type II, in scenarios of type I the asymmetry is generated in the sector which initiates both the CP-conserving and violating scatterings in the first place. This dual role is highlighted for the first time in this paper.

We shall now demonstrate the above scenarios through illustrative examples. In Sec.~\ref{sec:sec2} we discuss a leptogenesis scenario of type A-II, in which both the CP-conserving and violating processes are decays, and the asymmetry is generated in the daughter sector. In Sec.~\ref{sec:sec3} we discuss two scenarios of asymmetric dark matter production from scattering of type B-I, in which both the processes are scatterings, and the asymmetry is generated in the mother sector. We summarize our findings in Sec.~\ref{sec:sec4}.

\section{CP-violating and conserving decays and daughter sector asymmetry}
\label{sec:sec2}
We first discuss an illustrative model which belongs to the type A-II in the classification described in the introduction, i.e., both the CP-conserving and violating processes are two-body decays, and the asymmetry is generated in the daughter sector. To this end, consider a leptogenesis model involving two standard model (SM) singlet heavy Majorana neutrinos $N_1$ and $N_2$, with their mass values satisfying $M_{N_1}>M_{N_2}$. For the cosmology of $N_1$, the following decays of $N_1$ to SM charged lepton ($\ell^\pm$), charged scalar boson ($H^\pm$), neutral scalar boson ($h$) and $N_2$ are important:
\begin{align}
N_1 &\rightarrow \ell^{\pm} H^{\mp}~~~~~{\rm (CP-violating)} \nonumber \\
N_1 &\rightarrow N_2 h ~~~~~{\rm (CP-conserving)}. \nonumber
\end{align}

In order to realize these processes obeying the SM gauge symmetries, we need to extend the SM field content in the scalar sector. In addition to the dominantly SM-like scalar doublet $\Phi_1$ which gives mass to the SM fermions, we introduce a second Higgs doublet $\Phi_2$, and an SM singlet scalar field $S$. The  relevant interaction Lagrangian terms before electroweak symmetry breaking include the following:
\begin{equation}
\mathcal{L}_{\rm int} \supset -\frac{g}{2} \overline{N_1} N_2 S - y_i \overline{L} \tilde{\Phi}_1 N_i -  \mu_i |\Phi_i|^2 S -  \frac{\lambda^S_i}{2}  |\Phi_i|^2 S^2 - \lambda (\Phi_1^\dagger \Phi_1) (\Phi_2^\dagger \Phi_2)
\end{equation}
where $\tilde{\Phi}_1=i\sigma_2 \Phi_1^*$. Here, we have written the interaction terms involving the SM singlet Majorana neutrinos using the mass eigenstates $N_1$ and $N_2$. In order to avoid tree-level flavour-changing neutral currents, we have also assumed an additional $Z_2$ symmetry, under which $\Phi_2 \rightarrow -\Phi_2$, that is softly broken by a $m_{12} \Phi_1^\dagger \Phi_2$ term, as in the type-I two Higgs doublet model~\cite{Branco}. This symmetry prohibits the $y^\prime_i \overline{L} \tilde{\Phi}_2 N_i$ terms. After electroweak symmetry breaking, the singlet scalar mixes with the neutral components of the scalar doublets, giving rise to the decay mode $ N_1 \rightarrow N_2 h $, while the mixing of the charged scalars leads to the decay $N_1 \rightarrow \ell^{\pm} H^{\mp}$ through the Yukawa interaction.

Such a model was proposed in Ref.~\cite{Bhattacharya:2011sy} as a low-scale model for leptogenesis that utilizes the quartic scalar couplings~\cite{Kayser:2010fc}. In particular, the lepton number conserving decay mode $N_1 \rightarrow N_2 h$ helps in satisfying the requirements of generating a non-zero CP-violation at the one-loop level, being consistent with the Nanopoulous-Weinberg theorem~\cite{Nanopoulos:1979gx, Bhattacharya:2011sy}. However, the role of the CP-conserving decay mode $N_1 \rightarrow N_2 h$ in the cosmology of $N_1$ --- and therefore in the generated lepton asymmetry --- has not been studied earlier, and is the focus of the present study.

The Boltzmann equations determining the number density of $N_1$, and the lepton asymmetry produced are given by\footnote{Since $M_{N_1}>M_{N_2}$, during the out-of-equilibrium decay of $N_1$ at a temperature $T\simeq M_{N_1}$, the $N_2$ particles are in thermal equilibrium through rapid scatterings in the thermal bath, and hence follow a Maxwell-Boltzmann distribution. Furthermore, for simplicity, we assume that $N_2$ decays after the freeze-out of $N_1$ decay do not wash-out the generated lepton asymmetry, which can be ensured, for example, with $M_{N_2}<M_{H^{\pm}}$.}
\begin{align}
\dfrac{dY_{N_1}}{dx} &\simeq -\dfrac{\expval{\Gamma}_0+\expval{\Gamma}_A}{Hx}\, (Y_{N_1}-Y_{N_1,0}) \nonumber\\
\dfrac{dY_L}{dx} &\simeq -\dfrac{\epsilon\expval{\Gamma}_0}{Hx} \,(Y_{N_1}-Y_{N_1,0}).
\label{eq:Boltz_Decay}
\end{align}
Here, $Y_i = n_i/s$ is the yield of the species $i$, with $n_i$ being its number density and $s$ the entropy density in the radiation bath. We also have $Y_L = Y_{\ell^-}-Y_{\ell^+}$ as the lepton asymmetry produced, and $Y_{N_1,0}$ the equilibrium yield of $N_1$. Here it is assumed that at high temperatures $T>M_{N_1}$, $N_1$ achieves a thermal distribution through rapid scattering processes in the thermal plasma. The thermally averaged symmetric decay width is given by $\expval{\Gamma}_0$, where, $\Gamma_0=\Gamma(N_1 \rightarrow \ell^+ H^-)+\Gamma(N_1 \rightarrow \ell^- H^+)$, and $\Gamma_A$ is the decay width of the CP-conserving mode $N_1 \rightarrow N_2 h$. Finally, the CP-violation parameter is defined as 
\begin{equation}
\epsilon = \frac{|M(N_1 \rightarrow \ell^- H^+)|^2-|M(N_1 \rightarrow \ell^+ H^-)|^2}{|M(N_1 \rightarrow \ell^- H^+)|^2+|M(N_1 \rightarrow \ell^+ H^-)|^2}
\label{eq:eps}
\end{equation}
Here, $|M|^2$ denotes the matrix element squared for the corresponding process. In writing the Boltzmann equation for $Y_L$, we have made the approximation that at the epoch of the generation of the asymmetry, the two-body scattering reactions have decoupled, which is as expected due to their lower rates. We have also dropped terms in the right hand side proportional to $Y_L$, as at the epoch when $Y_L$ is being generated, those are subdominant.

\begin{figure}[htb!]
\centering
\includegraphics[scale=0.55]{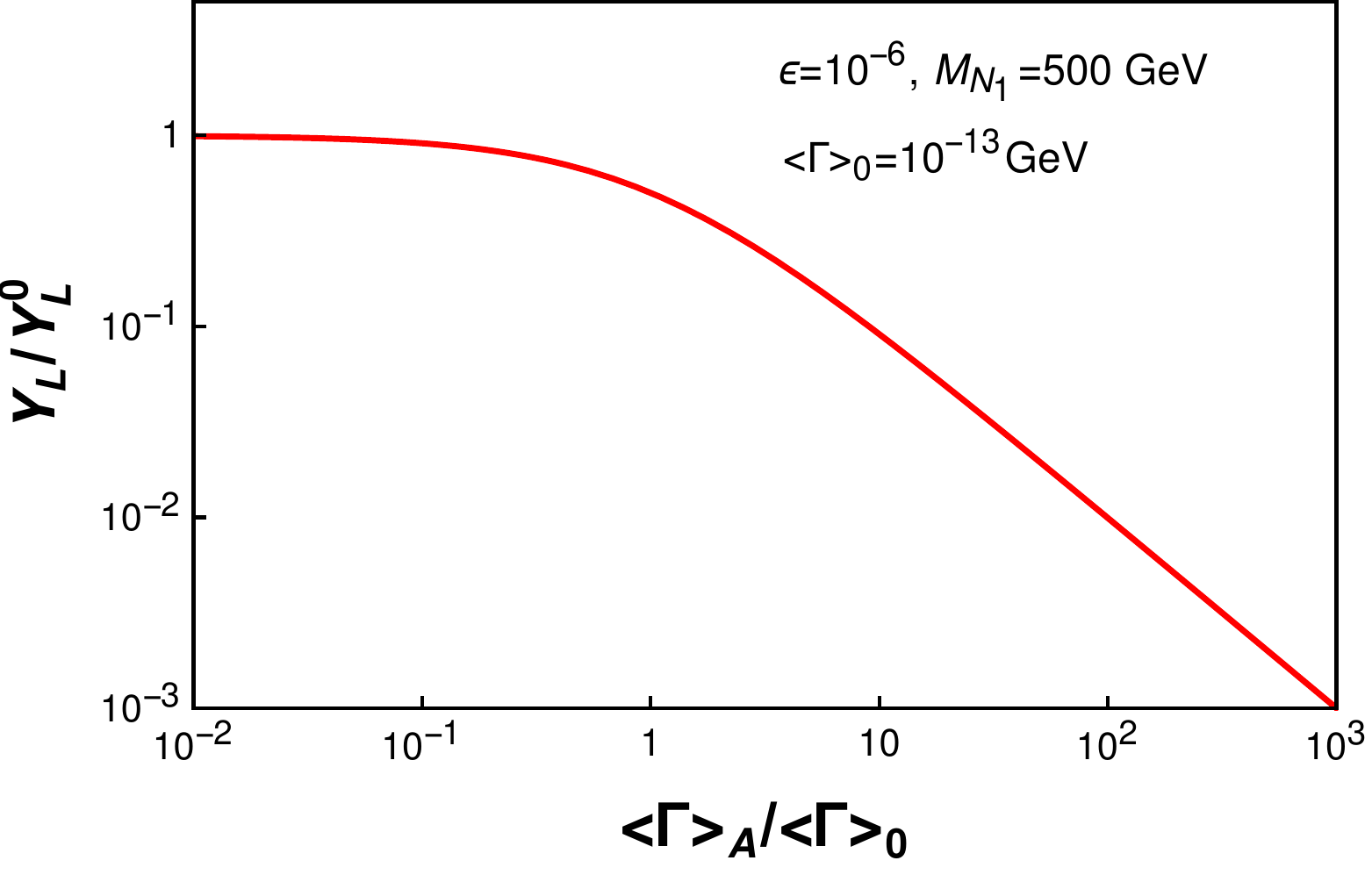}
\caption{\small{\it{The dependence of the lepton asymmetry $Y_L$ (scaled by $Y_{L,0})$, as a function of the ratio of CP-conserving and violating decay widths, $\expval{\Gamma}_A/\expval{\Gamma}_0$. See text for details on the choice of the other parameters.}}}
\label{Fig:lepto}
\end{figure}

We solve Eqs.~\ref{eq:Boltz_Decay} for the parameter choices $M_{N_1}=500$ GeV, $\epsilon=10^{-6}$ and $\Gamma_0=10^{-13}$ GeV, where the parameters are chosen such that the required baryon asymmetry of the Universe may be reproduced. We show the dependence of $Y_L$ (scaled by $Y_{L}^0$, which is the asymmetric yield with $\Gamma_A=0$) as a function of $\expval{\Gamma}_A/\expval{\Gamma}_0$ in Fig.~\ref{Fig:lepto}. As we can clearly see from this figure, the lepton asymmetry $Y_L$ changes by one order of magnitude if $\Gamma_A \sim 10 \Gamma_0$, compared to the scenario where $\Gamma_A$ is negligible. This simply stems from the fact that the longer the CP-conserving process is in equilibrium, the smaller the number density of the mother particle $N_1$, and hence the lepton asymmetry generated from its decays. 

\begin{figure}[htb!]
\centering
\includegraphics[scale=0.55]{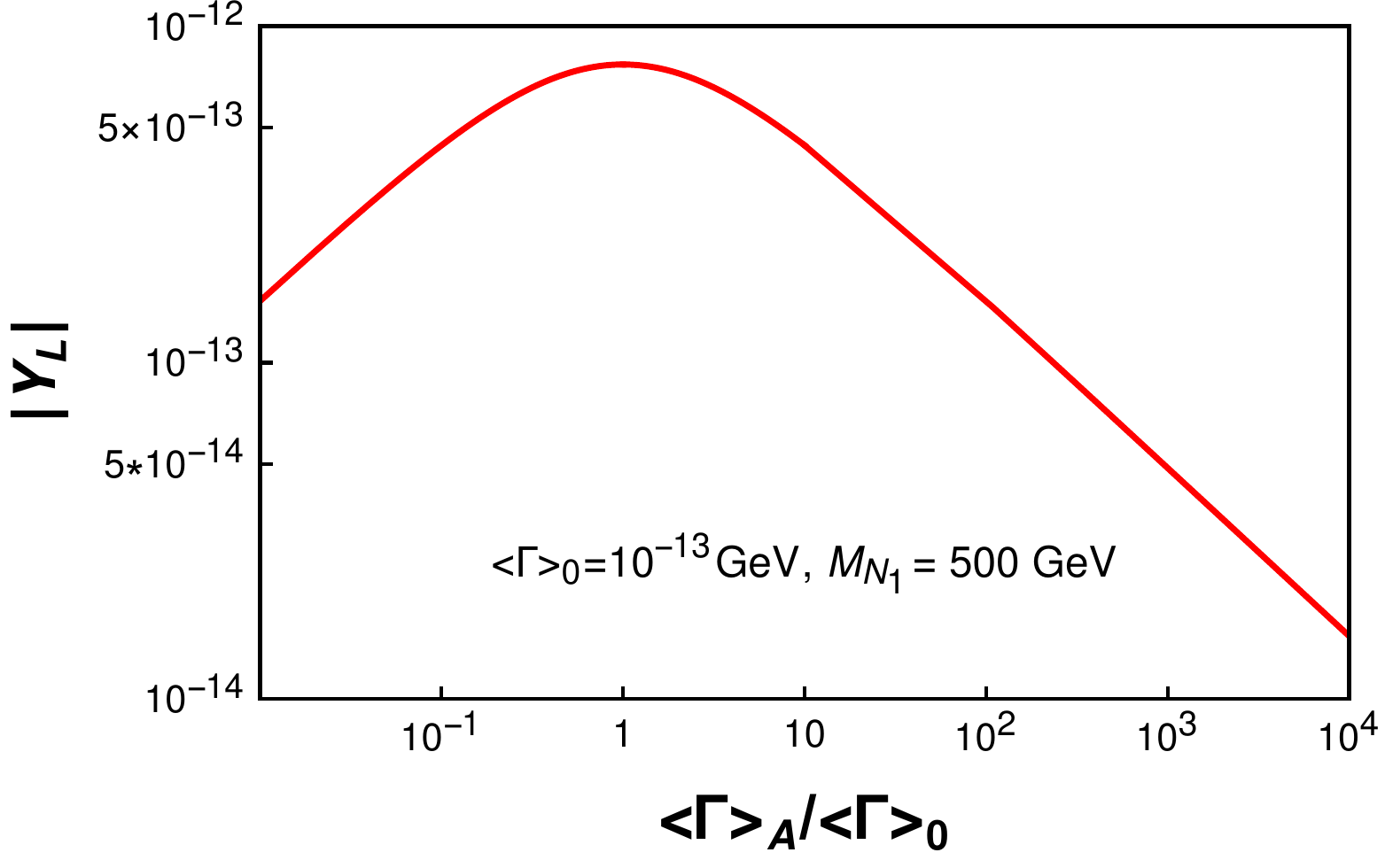}
\caption{\small{\it{The dependence of the lepton asymmetry $|Y_L|$ in the model for low-scale leptogenesis, as a function of the ratio of CP-conserving and violating decay widths, $\expval{\Gamma}_A/\expval{\Gamma}_0$. In this model, the CP-violation parameter $\epsilon$ can be expressed in terms of $\Gamma_A$, for fixed values of the scalar quartic coupling $\lambda$ and $\tan \beta$. See text for details.}}}
\label{Fig:lepto2}
\end{figure}

One can simplify the analysis further in the particular low-scale leptogenesis model discussed above. In this scenario, we can express the CP-violation parameter $\epsilon$ defined in Eq.~\ref{eq:eps}, stemming from the interference of the tree and loop level amplitudes, as follows:
\begin{equation}
\epsilon \simeq -2\lambda\, v\, \sin\beta \,\tilde{g}\, {\rm Im} (L).
\label{eq:lepto_eps}
\end{equation}
Here, $v=246$ GeV is the electroweak symmetry breaking scale and $\tan \beta = v_1/v_2$ is the ratio of the vacuum expectation values of the neutral CP-even components of the two Higgs doublets $\Phi_1$ and $\Phi_2$. We have also defined $\tilde{g}=g\, \sin\alpha$, where $\alpha$ is the mixing angle of the scalar singlet $S$ with the lighter CP-even Higgs boson. For simplicity, here we have assumed that $S$ dominantly mixes with the SM-like lighter Higgs state. Finally, ${\rm Im} (L)$ is the imaginary part of the loop-factor, which is found to be:
\begin{equation}
{\rm Im} (L) \simeq -\frac{M_{N_2}}{8\pi M^2_{N_1}}\frac{1}{1-\xi}\log \left(\frac{1}{\xi} \right),
\end{equation}
where, $\xi = \left(M_{H^\pm}/M_{N_1}    \right)^2$. For example, with $M_{N_1}=500$ GeV, $M_{N_2}=300$ GeV and $M_{H^\pm}=350$ GeV, we obtain ${\rm Im} (L) \simeq -6.7 \times 10^{-5}$. 

Since the width of the CP-conserving decay $N_1 \rightarrow N_2 h$ is $\Gamma_A = \frac{\tilde{g}^2}{8\pi}M_{N_1}$, we can trade the coupling $\tilde{g}$ in Eq.~\ref{eq:lepto_eps} with the square-root of the CP-conserving decay width $\Gamma_A$. In addition, if we fix the scalar quartic coupling $\lambda$ and $\tan \beta$, $\epsilon$ is determined in terms of $\Gamma_A$. The residual dependence of the lepton asymmetry $|Y_L|$ in this model as a function of $\expval{\Gamma}_A/\expval{\Gamma}_0$ is shown in Fig.~\ref{Fig:lepto2}, where $\expval{\Gamma}_0$ is chosen to be $10^{-13}$ GeV, $\sin \beta \sim \mathcal{O}(1)$ and the quartic coupling $\lambda \sim \mathcal{O}(0.1)$. As we can see from this figure, for $\expval{\Gamma}_A<\expval{\Gamma}_0$, the out-of-equilibrium condition for $N_1$ is determined by the CP-violating decay. Therefore, 
$|Y_L|$ increases with increasing $\expval{\Gamma}_A$, simply because the CP-violation $\epsilon$ increases as $\sqrt{\Gamma_A}$, for a fixed scalar quartic. On the other hand, for $\expval{\Gamma}_A > \expval{\Gamma}_0$, $|Y_L|$ decreases rapidly with increasing 
$\expval{\Gamma}_A$, as $N_1$ remains in equilibrium for a longer period, thereby leading to a reduction in its number density. We would like to emphasize that $\Gamma_A >>  \Gamma_0$ is easily achieved in the model described above for natural choices of the parameters.

Thus, CP-conserving processes, when at the same footing as the CP-violating ones (in this case both of them being decays of $N_1$), can play a major role in deciding the lepton asymmetry production in the early Universe. As discussed in the introduction, this is in contrast to the role of such processes observed earlier in the leptogenesis models from different see-saw mechanisms, as in those cases the CP-conserving modes played a subdominant role, mostly because they were scattering reactions, whereas the violating ones were decays, and hence of a higher rate.

\section{CP-violating and conserving scatterings and mother sector asymmetry}
\label{sec:sec3}
We shall now discuss two scenarios of asymmetric dark matter production from scattering. These two examples belong to the type B-I explained in the Introduction, in which both the CP-violating and conserving processes are scatterings, and the asymmetry is generated in the mother sector. We shall see that due to the latter feature, there is a dual role played by the CP-conserving processes. The ADM models under discussion have been proposed by the present authors in Ref.~\cite{GGM_ADM_Z2}, which discusses ADM from scattering, and in Ref.~\cite{Ghosh:2020lma} on ADM from semi-annihilation. 

Both the ADM models share some common features. There is a complex scalar DM $\chi$, which is stabilized by a $Z_N$ symmetry, interacting with itself and a $Z_N$ even real scalar $\phi$. The scalar $\phi$, taken to be lighter than $\chi$,  can mix with or decay to SM states, thereby maintaining kinetic equilibrium of the DM sector with the SM bath.\footnote{In principle, we can directly couple $\chi$ with the SM Higgs doublet $H$ through the $\lambda_{\chi H} |\chi|^2 |H|^2$ term. However, DM direct detection probes, as well as the constraints on the invisible decay width of the Higgs lead to a strong bound on such an interaction. This bound, in turn, makes it difficult to obtain the necessary thermal relic density for a large range of DM mass. Therefore, as a simplifying approximation, we have set $\lambda_{\chi H}\simeq 0$ for our discussion.} In both cases, the existence of a CP-conserving scattering $\chi+\chi^\dagger \rightarrow  \phi+\phi$ immediately follows, which is the focus of this discussion. In the ADM model from scattering, the stabilizing symmetry is $Z_2$, while for the ADM from semi-annihilation scenario, it is $Z_3$. Let us briefly discuss each model in the following. 

\subsection{Asymmetric DM from scattering}
\label{sec:ADM_1}
In this case, we have the following interaction Lagrangian, consistent with the $Z_2$ symmetry:
\begin{align}
-\mathcal{L_{\rm int}}  \supset  \mu  \chi^\dagger \chi  \phi+ \left(\frac{\mu_1}{2} \chi^2 \phi+ {\rm h.c.} \right)  + \frac{\lambda_1}{4} \left(\chi^\dagger \chi \right)^2 +   \left(\frac{\lambda_2}{4!}\chi^4 +{\rm h.c.} \right)+\left(\frac{\lambda_3}{4}\chi^2 \phi^2 +{\rm h.c.} \right)\nonumber\\
+\left(\frac{\lambda_4}{3!}\chi^3\chi^{\dagger} +{\rm h.c.} \right)
+ \frac{\lambda_5}{2} \phi^2 \chi^\dagger \chi  +  \frac{\mu_\phi}{3!} \phi^3 +  \frac{\lambda_\phi}{4!} \phi^4 + \frac{\lambda_{\phi H}}{2} \phi^2 |H|^2 + \mu_{\phi H} \phi |H|^2 .\,\,\,
\label{eq:lag_z2}
\end{align}
Here, $H$ is the SM Higgs doublet. The neutral particle $\phi$ can mix with the Higgs boson after electroweak symmetry breaking, thus inducing direct detection signals for DM through the trilinear interactions with couplings $\mu$ and $\mu_1$. To evade the current direct detection bounds, we have set $\mu\simeq 0$ and $\mu_1\simeq 0$, rendering only the contact interactions to be relevant for our discussion.

Several relevant processes for the DM cosmology resulting from these interactions are listed in Table~\ref{Tab:tab1}, along with their properties and the notation used to denote the different thermally averaged symmetric and asymmetric reaction rates. The asymmetric and symmetric thermal averages are defined with and without $\epsilon_f$ for each process (the definition of $\epsilon_f$  is analogous to Eq.~\ref{eq:eps}, which, for scatterings, is an explicit function of the particle momenta $p_i$), where $\expval{\epsilon \sigma v}_{f}$ can be written as:
\begin{equation}
 \expval{\epsilon \sigma v}_{f} = \dfrac{\int \prod^{4}_{i=1} \frac{d^3 p_i}{(2\pi)^3 2 E_{p_i}}  (2 \pi)^4 \delta^{(4)}(p_1+p_2-p_3-p_4) \,\epsilon_f(p_i) |M_0|^2_f f_0(p_1)f_0(p_2)}{\int \dfrac{d^3 p_1}{(2\pi)^3} \dfrac{d^3 p_2}{(2\pi)^3} f_0(p_1)f_0(p_2)} \hspace{0.5cm},
\label{eq:cross}
\end{equation}
with   $|M_0|^2_f = |M|^2_{\chi\chi\rightarrow f}+|M|^2_{\chi^{\dagger}\chi^{\dagger}\rightarrow f^{\dagger}} $ and $f_0(p)$ being the equilibrium distribution function.
\begin{table}[htb!]
\begin{center}
\begin{tabular}{|c|c|c|c|c|}
\hline
Process & CP & DM Number & Rate & Rate \\
              &      & Violation  & (Symmetric) & (Asymmetric) \\
\hline
$\chi+\chi \rightarrow \chi^{\dagger} + \chi^{\dagger}$ & violating & 4 units & $\expval{\sigma v}_1$ &  $\expval{\epsilon \sigma v}_1$\\
\hline
$\chi+\chi \rightarrow \chi + \chi^{\dagger}$ & violating & 2 units & $\expval{\sigma v}_2$  &  $\expval{\epsilon \sigma v}_2$ \\
\hline
$\chi+\chi \rightarrow \phi+\phi$ & violating & 2 units & $\expval{\sigma v}_3$ & $\expval{\epsilon \sigma v}_3$  \\
\hline
$\chi+\chi^\dagger \rightarrow  \phi+\phi$ & conserving & 0 units & $\expval{\sigma v}_A$ & NA \\
\hline
\end{tabular}
\end{center}
\caption{\small{\it{Relevant processes for DM cosmology for the ADM from scattering scenario, their properties, and notation used to denote the corresponding thermally averaged symmetric and asymmetric reaction rates.}}}
\label{Tab:tab1}
\end{table}%

With the above reactions in the thermal plasma, the evolution equations for the symmetric ($Y_S=Y_\chi+Y_{\chi^\dagger}$) and asymmetric yields ($Y_{\Delta \chi}=Y_\chi -Y_{\chi^\dagger}$) in this scenario are given as follows:

\begin{align}
\label{asym}
\dfrac{dY_S}{dx}&=-\dfrac{s}{2Hx}\left[\expval{\sigma v}_A\left(Y^2_S-Y^2_{\Delta \chi}-4Y^2_0\right)+\expval{\sigma v}_3\bigg(\dfrac{Y^2_S+Y^2_{\Delta \chi}-4Y^2_0}{2}\bigg)-\expval{\epsilon\sigma v}_S Y_S Y_{\Delta \chi}\right]\nonumber\\
\dfrac{dY_{\Delta \chi}}{dx}&=-\dfrac{s}{2Hx}\left[\expval{\epsilon\sigma v}_S\left(\dfrac{Y^2_S-4Y^2_0}{2}\right)+\expval{\epsilon\sigma v}_D \dfrac{Y^2_{\Delta \chi}}{2}+\expval{\sigma v}_{all}\,Y_S Y_{\Delta \chi}\right].
\end{align} 
Here, we have defined, $\expval{\epsilon\sigma v}_S =\expval{\epsilon \sigma v}_1+\expval{\epsilon \sigma v}_2 $, $\expval{\epsilon\sigma v}_D=\expval{\epsilon \sigma v}_1-\expval{\epsilon \sigma v}_2$ and $\expval{\sigma v}_{all}=2\expval{\sigma v}_1+\expval{\sigma v}_2+\expval{\sigma v}_3$. In writing these equations, we have used the unitarity sum rule relating the $\expval{\epsilon \sigma v}_i$'s, namely, $\expval{\epsilon \sigma v}_1+\expval{\epsilon \sigma v}_2+\expval{\epsilon \sigma v}_3=0$, and have eliminated $\expval{\epsilon \sigma v}_3$, in terms of the other two asymmetric rates~\cite{Kolb:1979qa}.

\begin{figure}[htb!]
\centering
\includegraphics[scale=0.55]{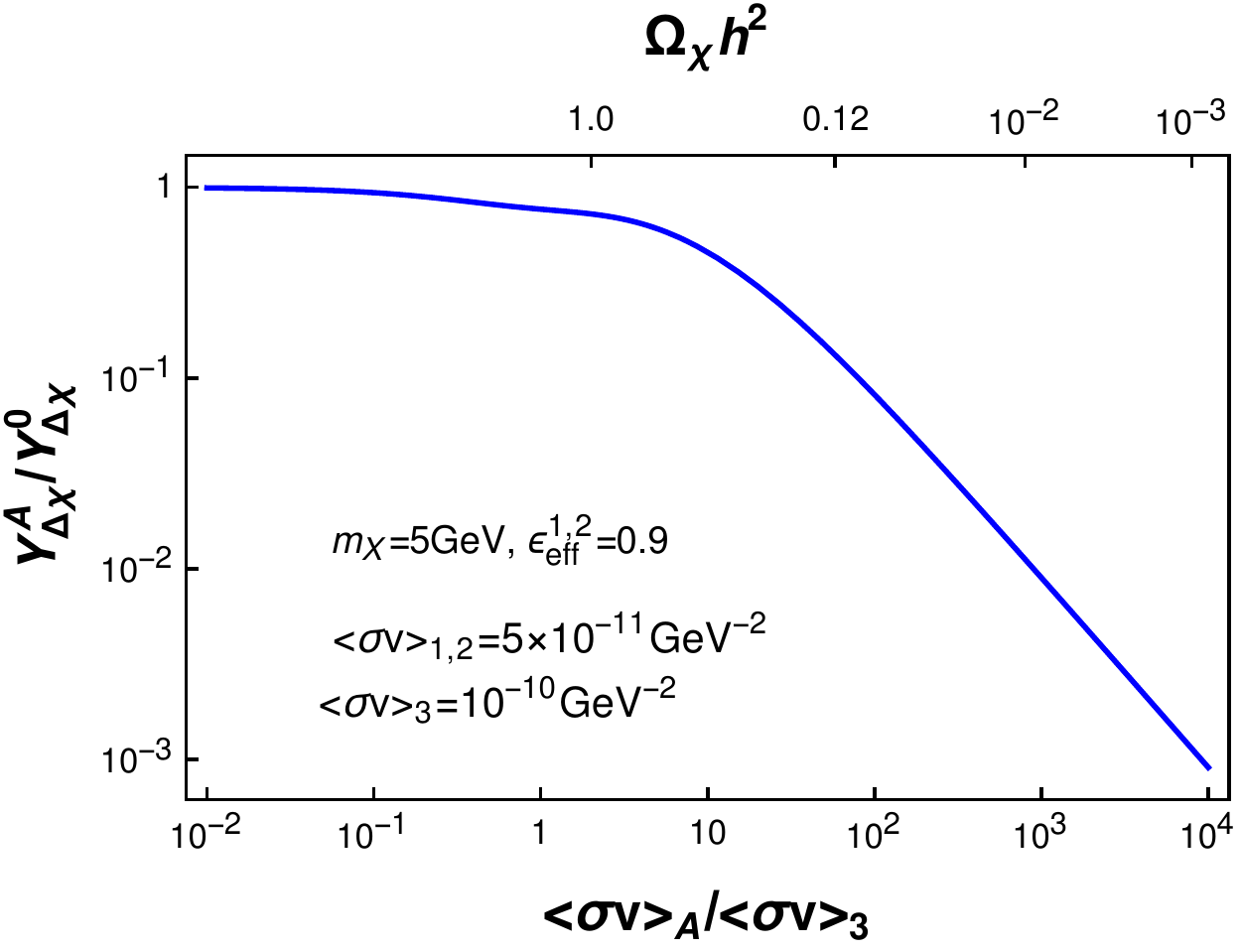}
\includegraphics[scale=0.5]{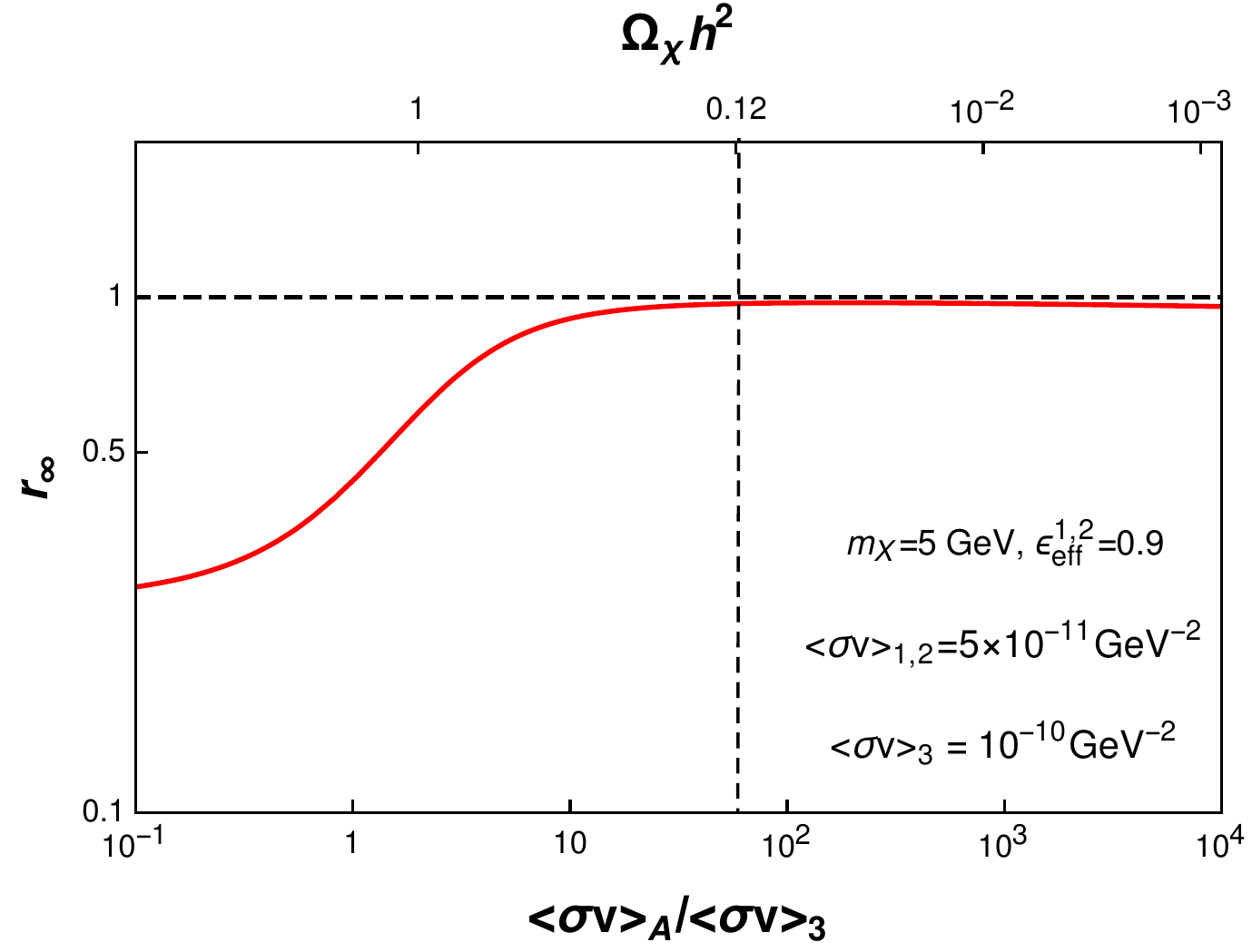}
\caption{\small{\it{Left Panel:  The ratio, $Y_{\Delta \chi}^A/Y_{\Delta \chi}^0$, where $Y_{\Delta \chi}^A$ and $Y_{\Delta \chi}^0$ are the asymptotic values of the yield $Y_{\Delta \chi}$, with and without the CP-conserving annihilation process, respectively, as a function of $\expval{\sigma v}_A/\expval{\sigma v}_3$. Also shown are the corresponding values of the DM relic abundance $\Omega_\chi h^2$. Right Panel: The final particle-antiparticle asymmetry parameter $r_\infty=\lvert Y_{\Delta\chi} \rvert / Y_S$, as a function of $\expval{\sigma v}_A/\expval{\sigma v}_3$. See text for details. Both the figures are for the asymmetric DM model from scattering discussed in Sec.~\ref{sec:ADM_1}.}}}
\label{FigZ2}
\end{figure}

Solving Eqs.~\ref{asym} numerically we can determine the impact of the CP-conserving annihilation process in modifying the asymmetric yield. To this end, we show the ratio $Y_{\Delta \chi}^A/Y_{\Delta \chi}^0$, where $Y_{\Delta \chi}^A$ and $Y_{\Delta \chi}^0$ are the asymptotic values of the yield $Y_{\Delta \chi}$, with and without the CP-conserving annihilation process, respectively. We show the yield ratio as a function of $\expval{\sigma v}_A/\expval{\sigma v}_3$ in Fig.~\ref{FigZ2} (left panel), which is the ratio of the CP-conserving and CP-violating annihilations that control the out-of-equilibrium number densities. Here, we have kept the symmetric and asymmetric reaction rates for all the CP-violating processes, and the DM mass, as fixed. For 
$\expval{\sigma v}_A \gtrsim \expval{\sigma v}_3$, we see that $Y_{\Delta \chi}$ reduces significantly, by a factor of $10$ for $\expval{\sigma v}_A/\expval{\sigma v}_3 \sim 75$. This feature can also be seen by solving Eqs.~\ref{asym} analytically near freeze-out, with the following result, as discussed in detail in Ref.~\cite{GGM_ADM_Z2}:
\begin{equation}
|Y_{\Delta\chi}(x)|= \dfrac{2 H x}{s}\,\dfrac{\expval{\epsilon\sigma v}_S}{\left(2\expval{\sigma v}_A + \expval{\sigma v}_3 \right)\expval{\sigma v}_{all}+\expval{\epsilon \sigma v}^2_S} .
\label{analytic}
\end{equation}
Thus we see that as the CP-conserving reaction rate $\expval{\sigma v}_A $ increases, $Y_{\Delta\chi}(x)$ is reduced, which, in turn, tends to reduce the observed particle-antiparticle asymmetry as well. However, there is a second role of these CP-conserving pair-annihilations, which appears in the epoch after the CP-violating reactions are frozen out. The symmetric component of the DM is then subsequently eliminated by the $\chi+\chi^\dagger \rightarrow  \phi+\phi$ reaction, thus modifying the ratio of particle anti-particle yields at the late epochs.

In order to demonstrate the second effect, it is useful to define the final particle-antiparticle asymmetry parameter $r_\infty$ as follows: 
\begin{equation}
r_\infty =\lvert Y_{\Delta\chi} \rvert / Y_S
\end{equation}
where the asymptotic values of the yields with $x=M_{N_1}/T \rightarrow \infty$ have been used. Clearly, $0 \leq r_\infty \leq 1$, where $r_\infty=0$ corresponds to the completely symmetric limit, in which the asymptotic yields of the DM and anti-DM are the same, while $r_\infty=1$ corresponds to the completely asymmetric limit, in which only either the DM or the anti-DM species survives. In the right panel of Fig.~\ref{FigZ2}, we show the $r_\infty$ parameter as a function of $\expval{\sigma v}_A/\expval{\sigma v}_3$, for fixed values of all the CP-violating rates. Similar to the example for leptogenesis from decays, we have defined an \textit{effective} CP-violation parameter (independent of the particle momenta) for each channel as $\epsilon^f _{\rm eff}=\expval{\epsilon \sigma v}_f/\expval{\sigma v}_f$. We find that as we increase the value of $\expval{\sigma v}_A/\expval{\sigma v}_3$, $r_\infty \rightarrow 1$, and a completely asymmetric DM with the required relic abundance is obtained for $\expval{\sigma v}_A/\expval{\sigma v}_3 \sim 60$. This feature is obtained due to the second role of the CP-conserving process. However, the competition between the two roles is not visible in this figure, which becomes apparent in the second example of ADM production from semi-annihilation discussed in the next subsection.

We note in passing that for our choice of parameters, $\expval{\sigma v}_A/\expval{\sigma v}_3 \sim 60$ is also necessary to satisfy the DM relic abundance requirement of $\Omega_\chi h^2 =0.12$~\cite{Aghanim:2018eyx}. We have shown the value of $\Omega_\chi h^2$ for different values of $\expval{\sigma v}_A/\expval{\sigma v}_3 $ in Fig.~\ref{FigZ2} (both panels) as well. For $\expval{\sigma v}_A/\expval{\sigma v}_3 > 100$, the DM species is underabundant. This is again due to the fact that the CP-conserving process here plays a dual role of both changing  $Y_{\Delta \chi}$ at the earlier epoch, as well as reducing the symmetric DM component, and the latter eventually reduces the net relic density.

\subsection{Asymmetric DM from semi-annihilation}
\label{sec:ADM_2}
In the second example of type B-I, we consider the possibility of producing asymmetric dark matter from semi-annihilation~\cite{Ghosh:2020lma}, in which the dual role of the CP-conserving processes is demonstrated clearly. In this case, we have the following interaction Lagrangian, consistent with the $Z_3$ symmetry:
\begin{equation}
-\mathcal{L}_{\rm int} \supset \frac{1}{3!} \left(\mu \chi^3 + {\rm h.c.} \right) + \frac{1}{3!} \left(\lambda \chi^3 \phi + {\rm h.c.} \right) + \frac{\lambda_1}{4} \left(\chi^\dagger \chi \right)^2 +   \frac{\lambda_2}{2} \phi^2 \chi^\dagger \chi + \mu_1 \phi  \chi^\dagger \chi +  \frac{\mu_2}{3!} \phi^3 +  \frac{\lambda_3}{4!} \phi^4.
\label{eq:lag_z3}
\end{equation}
In addition, there will be interactions between $\phi$ and the SM Higgs doublet $H$, exactly as in Eq.~\ref{eq:lag_z2}. To realize the semi-annihilation scenario, for the necessary CP-violation through the interference of the tree and one-loop graphs, the two complex couplings $\mu$ and $\lambda$ are required in general, so that one complex phase remains after field redefinitions. We also note in passing that for the parameter region $\mu/m_\chi << 1$ and $\mu_1/m_\chi<<1$, the contact interactions play the dominant role, which we have assumed for illustration. 

These interactions lead to several relevant processes for the DM cosmology, which are shown in Table~\ref{Tab:tab2}, along with their properties and the notation used to denote the different thermally averaged reaction rates. 
\begin{table}[htb!]
\begin{center}
\begin{tabular}{|c|c|c|c|c|}
\hline
Process & CP & DM Number & Rate & Rate \\
              &      & Violation  & (Symmetric) & (Asymmetric) \\
\hline
$\chi+\chi \rightarrow \chi^{\dagger}+\phi$ & violating & 3 units & $\expval{\sigma v}_1$ &  $\expval{\epsilon \sigma v}_1$\\
\hline
$\chi+\chi \rightarrow \chi^{\dagger}+\phi+\phi$ & violating & 3 units & $\expval{\sigma v}_2$  &  $\expval{\epsilon \sigma v}_2$ \\
\hline
$\chi+\chi \rightarrow \chi^{\dagger}+\chi^{\dagger}+\chi$ & violating & 3 units & $\expval{\sigma v}_3$ & $\expval{\epsilon \sigma v}_3$  \\
\hline
$\chi+\chi^{\dagger} \rightarrow \phi +\phi$ & conserving & 0 units & $\expval{\sigma v}_A$ & NA \\
\hline
\end{tabular}
\end{center}
\caption{\small{\it{Relevant processes for DM cosmology for the ADM from semi-annihilation scenario, their properties, and notation used to denote the corresponding thermally averaged symmetric and asymmetric reaction rates.}}}
\label{Tab:tab2}
\end{table}%

The Boltzmann equations for the symmetric ($Y_S=Y_\chi+Y_{\chi^\dagger}$) and asymmetric yields ($Y_{\Delta \chi}=Y_\chi -Y_{\chi^\dagger}$) in this scenario are given by:

\begin{align}
\dfrac{dY_S}{dx} &= -\dfrac{s}{8Hx}\bigg[\expval{\sigma v}_S \left(Y^2_S+Y^2_{\Delta\chi}-2 Y_0 Y_S\right)+\expval{\sigma v}_3\left(Y^2_S\left(\dfrac{Y_S}{2Y_0}-1\right)-Y^2_{\Delta\chi}\left(\dfrac{Y_S}{2Y_0}+1\right)\right) \nonumber \\
 &+ 4\expval{\sigma v}_A\left(Y^2_S-Y^2_{\Delta \chi}-4Y^2_0\right)+\expval{\epsilon\sigma v}_{3}\dfrac{Y_{\Delta\chi}}{2Y_0}\left(Y^2_S-Y^2_{\Delta\chi}+ 4Y^2_0-8 Y_0 Y_S\right)\bigg ] \nonumber\\
\dfrac{dY_{\Delta \chi}}{dx} &= -\dfrac{3 s}{4Hx}\bigg[\expval{\sigma v}_S Y_{\Delta \chi} \left(Y_S+Y_0\right)+\expval{\sigma v}_3 Y_{\Delta \chi}\left(Y_S+\dfrac{Y^2_S-Y^2_{\Delta\chi}}{4Y_0}\right) \nonumber \\
&+ \expval{\epsilon\sigma v}_{3}\dfrac{Y_S}{4Y_0}(Y^2_S-Y^2_{\Delta\chi}-4Y^2_0)\bigg],
\label{boltz:Z3}
\end{align}  
where, we have defined $\expval{\sigma v}_S = \expval{\sigma v}_1+\expval{\sigma v}_2$. Here, the thermally averaged rate $\expval{\sigma v}_3$ has a significant $x-$dependence, given by  $\expval{\sigma v}_3 =\dfrac{1+2x}{\sqrt{\pi x}} e^{-x} (\sigma v_3)_{s}$, where $(\sigma v_3)_{s}$ is an $x-$independent  s-wave piece, and the exponential suppression factor stems from the
phase-space cost for producing an extra particle in the final state~\cite{Bhatia:2020itt}.

\begin{figure}[htb!]
\centering
\includegraphics[scale=0.55]{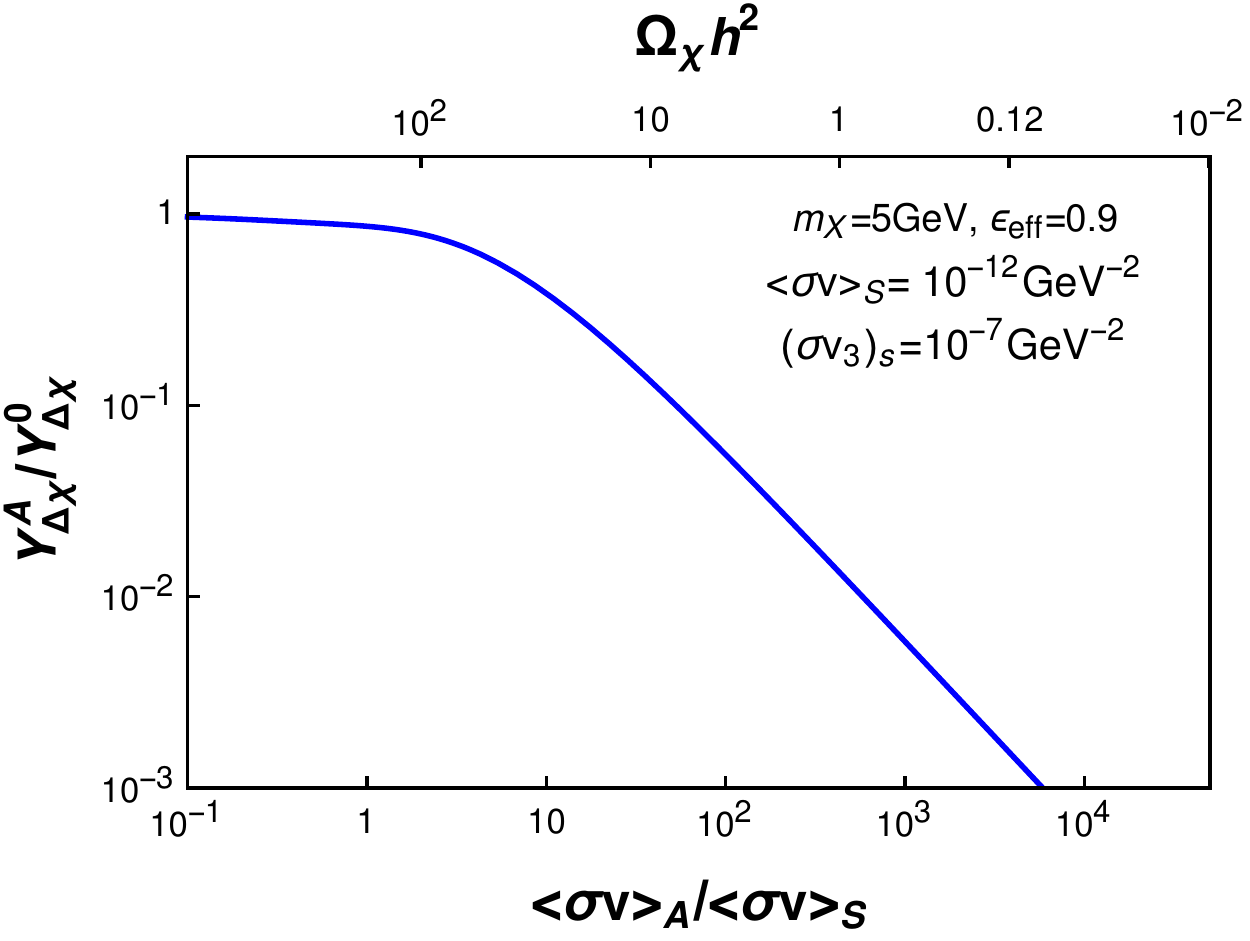}
\includegraphics[scale=0.55]{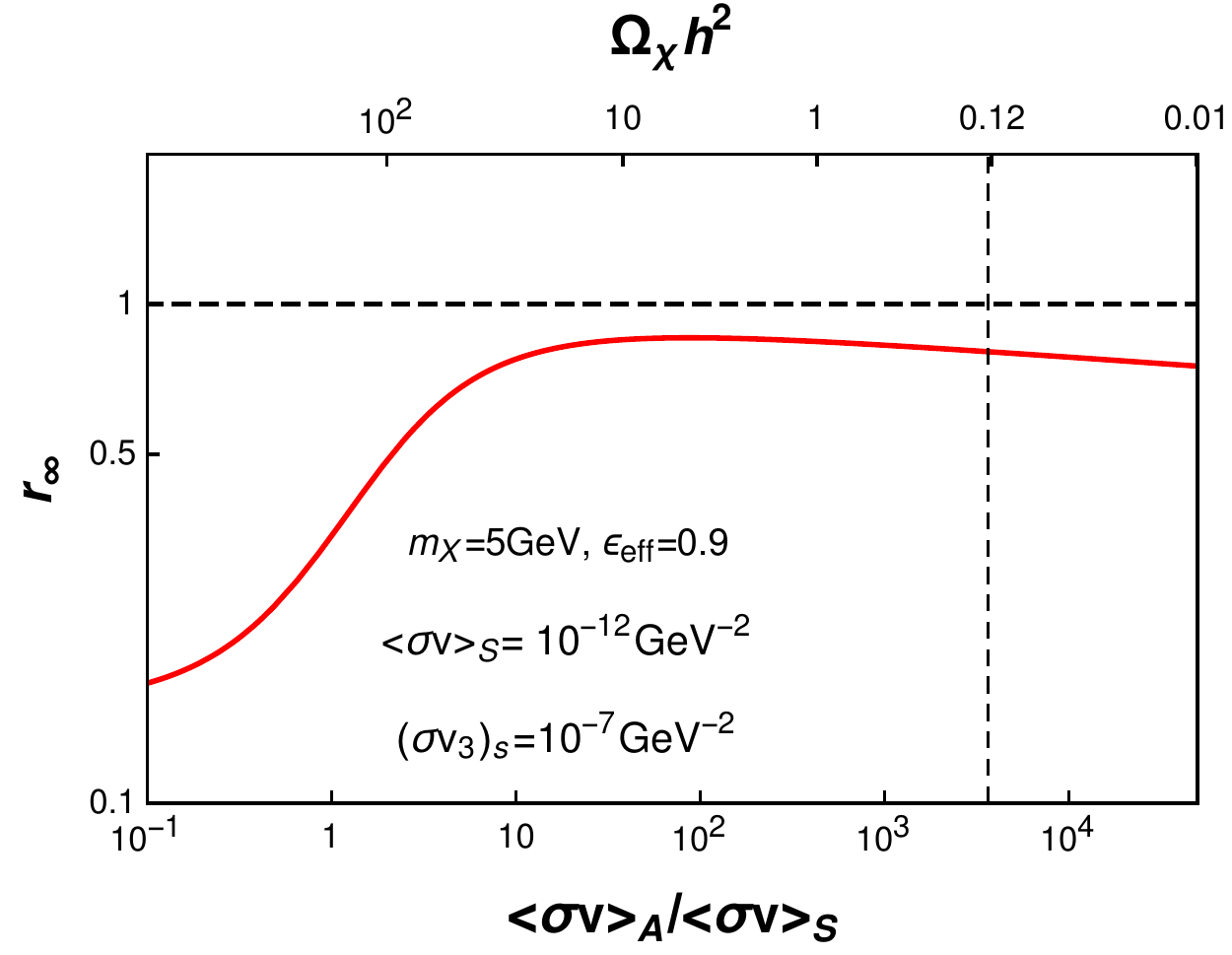}
\caption{\small{\it {Same as Fig.~\ref{FigZ2}, for the asymmetric DM  from semi-annihilation scenario discussed in Sec.~\ref{sec:ADM_2}. See text for details.}}}
\label{FigZ3}
\end{figure}

By solving Eqs.~\ref{boltz:Z3} numerically, we find the ratio of the asymmetric yields, as before, with and without the CP-conserving annihilation process, and show the results in the left panel of Fig.~\ref{FigZ3}, as a function of $\expval{\sigma v}_A /\expval{\sigma v}_S $. In the right panel of the same figure, we also show the final asymmetry parameter $r_\infty$ as defined in the previous subsection. 

Compared to the scenario in the previous subsection, the $r_\infty$ curve shows more features with the change in $\expval{\sigma v}_A /\expval{\sigma v}_S $. Initially, as we increase $\expval{\sigma v}_A$, $r_\infty$ approaches the completely asymmetric limit of $1$. However, on further increase of the ratio $\expval{\sigma v}_A /\expval{\sigma v}_S \gtrsim 100$, the value of $r_\infty$ is reduced instead. As before, there is a dual role played by the CP-conserving process -- that of reducing the value of $Y_{\Delta \chi}$ to begin with, and thus trying to reduce the asymmetry on the one hand, and at the same time of removing the symmetric component of the DM particles, and thus generating a competing effect to increase the final asymmetry parameter $r_\infty$ signifying the modification of the relative particle anti-particle yields. The initial increase of $r_\infty$ is observed as the second effect dominates for moderately large values of $\expval{\sigma v}_A /\expval{\sigma v}_S$. However, for very large values of $\expval{\sigma v}_A /\expval{\sigma v}_S$, the first effect dominates, reducing the $r_\infty$ to be lower than $1$. Thus the competition between the two effects at different epochs due to the same CP-conserving process is brought out clearly in this example of asymmetric dark matter production from semi-annihilation.

\section{Summary}
\label{sec:sec4}
To summarize, in this paper, we have revisited the role of CP-conserving processes in generating particle-antiparticle asymmetries in the early Universe -- either in the lepton sector for baryogenesis through leptogenesis, or in the dark matter sector. We have focussed on scenarios in which either both the CP-violating and conserving processes are decays, or both of them are scatterings, thus naturally being of comparable rates. For this reason, in the examples considered by us, the effect of the CP-conserving reactions is found to be highly significant. This is in contrast to the scenarios for leptogenesis in different see-saw models of neutrino mass, in which the role of CP-conserving reactions were explored earlier, and found to be mostly sub-dominant. In those scenarios, the primary source of CP-violation was a decay process, while the CP-conserving reaction was a scattering, and thus the latter process is generally Boltzmann suppressed. 

Within each scenario above, the asymmetry may be produced either in the mother sector, or in the daughter sector, leading to distinct effects of the CP-conserving processes in each case. As an example for a scenario in which both the CP-violating and conserving processes are decays and the asymmetry is generated in the daughter sector, we discussed a low-scale model for leptogenesis. It is shown that as the rate for the CP-conserving decays is increased compared to that of the CP-violating one, the lepton asymmetry yield proportionately decreases, and can vary by orders of magnitude for natural values of the model parameters. This is simply because, in such cases the CP-conserving decays remain longer in equilibrium, reducing the number density of the non-relativistic mother particles, and thereby reducing the generated lepton asymmetry. 

We then discussed two examples of asymmetric dark matter production in which both the CP-violating and conserving reactions are scatterings, and the asymmetry is generated in the mother sector. The latter feature leads to an interesting dual role played by the CP-conserving reactions. Initially, when the CP-violating reactions are active, the larger the rate of the CP-conserving scatterings, the smaller is the asymmetric yield of the particle-antiparticle system. However, subsequent to the freezing out of the CP-violating reactions, the CP-conserving pair annihilations remove the symmetric component of the DM, thereby enhancing the final asymmetry parameter, signifying the modification of the relative particle anti-particle yields. This dual role played by the same CP-conserving reaction leads to a novel competing effect observed in the ADM models. Thus CP-conserving processes can play qualitatively distinct roles in generating cosmological particle-antiparticle asymmetries, and can modify the asymmetric yields by orders of magnitude.


\section*{Acknowledgment}
The work of AG is partially supported by the RECAPP, Harish-Chandra Research Institute, and the work of DG is partially supported by CSIR, Government of India, under the NET JRF fellowship scheme with award file No. 09/080(1071)/2018-EMR-I, and in part by the Institute Fellowship provided by the Indian Association for the Cultivation of Science (IACS), Kolkata.

\end{document}